\documentclass[conference]{IEEEtran}
\ifCLASSINFOpdf

\else

\fi

\usepackage{url}
\usepackage[]{fancyhdr}
\usepackage{cite}
\usepackage{amsmath,amssymb,amsfonts}
\usepackage{algorithmic}
\usepackage{graphicx}
\usepackage{textcomp}
\usepackage{xcolor}
\usepackage{amsmath}
\usepackage{caption}
\newcommand{\changefont}{\fontsize{9}{9}\selectfont}
\IEEEoverridecommandlockouts
\begin{document}

\title{Adaptive and Robust Cross-Voltage-Level Power Flow Control of Active Distribution Networks\\} 

\author{\IEEEauthorblockN{Jannik Zwartscholten, Christian Rehtanz}
\IEEEauthorblockA{Institute of Energy Systems, Energy Efficiency and Energy Economics,\\TU Dortmund University, Dortmund, Germany\\
Email: jannik.zwartscholten@tu-dortmund.de}
}

\maketitle
\thispagestyle{fancy}
\pagestyle{fancy}

\begin{abstract}
The large-scale integration of Distributed Energy Resources (DERs) into the electric power system offers new opportunities to ensure stability. For example, Active Distribution Networks (ADNs) can be used in (sub-)transmission systems in the emergency state, as far as high robustness and performance of the ADN control are guaranteed. This paper presents an adaptive control system for ADN's cross-voltage-level power flow control. For this purpose, the gain scheduling approach is used. Furthermore, this work introduces a method for control parameter tuning. In order to validate the control parameter tuning, the adaptive control system is analyzed regarding robustness and performance using an exemplary medium voltage grid. In addition, the influence of uncertainties is examined. Finally, the operation of the adaptive control system is demonstrated by performing time-domain simulations.
\end{abstract}


\IEEEpeerreviewmaketitle

\section{Introduction}
\label{intro}

With the energy transition, the transformation of the electrical energy system continues steadily. The shift in power generation results in less controllable units in the transmission system and more controllable units in the distribution system \cite{entso2020TYNDP}. The installed capacity of renewable energies will exceed multiple times the peak load demand \cite{entso2020TYNDPSR}. For this reason, there will be more situations in which Distributed Energy Resources (DERs) together with controllable grid assets must guarantee the system's stability in the future. Therefore system operation must be adapted to the new conditions concerning ancillary services and system stability. 
 
In order to counteract the lack of control options in the transmission grid, one potential approach is to control DERs in clusters to provide ancillary services for the transmission grid or a higher voltage level in general. In the frame of Active Distribution Networks (ADNs), all DERs connected to the same distribution grid could be defined as such a cluster \cite{gonzalez2017dynamic,gonzalez2018flex}. In \cite{zwartscholten2018hierarchical,gonzalez2017dynamic,robitzky2017impact,greve2015simulation}, ADN concepts with a Cross-Voltage-Level Power Flow Control (CPFC) are described. In these approaches, a PI-based controller is able to control the ADNs' power flow at the interconnection point to the next higher voltage level.  Thereby, DERs in the ADN are the actuators of the control system. 

In the mentioned publications, it is assumed that the number of DERs available to the control system is constant. However, the availability of DERs can fluctuate over time for various reasons. For example, DERs may not be able to participate in the distribution grid control due to a lack of wind or the absence of solar radiation. In addition, discharged energy storage devices can lose their full flexibility potential. Also, Information and Communication Technology (ICT) errors or maintenance can cause DERs to be unavailable to the control system.
For this reason, there is a need for an adaptive control system, which adjusts the control parameters based on the available DERs. Without an adaptation mechanism, the control parameters would have to be designed based on the worst available DER combinations in terms of stability. Therefore, static control parameters would result in low performance of the control system. In emergency situations, the activation time of a countermeasure can strongly influence the outcome of a fault \cite{robitzky2018mod}. Hence, a high control performance of the ADN can be essential to maintain power system stability.
    
Furthermore, \cite{zwartscholten2018hierarchical,gonzalez2017dynamic,robitzky2017impact,greve2015simulation} assume that the dynamic behavior of the active and reactive setpoint tracking of DERs is known. However, Distribution System Operators (DSOs) have little to no information about the dynamic behavior of the DERs connected to their distribution grids. Creating detailed dynamic models of all DERs participating in the control loop requires a disproportionate effort. Therefore, the control system has to be designed by using simplified models of DERs. In doing so, model uncertainties must be taken into account.

This work extends the control concept from \cite{zwartscholten2018hierarchical} to an adaptive and robust control system. For this purpose, an adaption mechanism based on the gain scheduling method is added to the control structure. In addition, this work presents a method for determining the control parameters of the gain scheduling controller. In order to determine the parameters, the DERs are modeled as simplified linear systems plus time delay. Because the real dynamic behavior of the DERs differs from the simplified models, an uncertainty analysis is carried out. In addition, the robustness and performance of the control system are considered. 

In order to demonstrate the proposed control design, time-domain simulations are presented using the SimBench 20\,kV rural benchmark grid and detailed DER models. For simplification, line overloading and voltage band violations are not considered in this work. Besides that, the focus is on the control of active power, whereas the concept can also be used for the control of reactive power.

The paper is structured as follows. First, in Section \ref{CPFCADN}, the state of the art in research of ADN control is presented. Section \ref{adapcontrol} describes the development and structure as well as the control parameter tuning of the adaptive control system. Afterward, a simplified Linear Time-Invariant (LTI) based model and a detailed simulation model used in the investigations of this work are depicted in Section \ref{modelling}. Section \ref{results} presents the analysis of robustness, performance, and the influence of uncertainties, as well as time-domain simulations. Finally, in Section \ref{conclusion}, a conclusion is drawn discussing the newly developed approach and providing an outlook on future research in the field of ADN control.

\section{Cross-Voltage-Level Power Flow Control in Active Distribution Networks}
\label{CPFCADN}

\begin{figure}[h]
	\centering
	\includegraphics[scale=0.5]{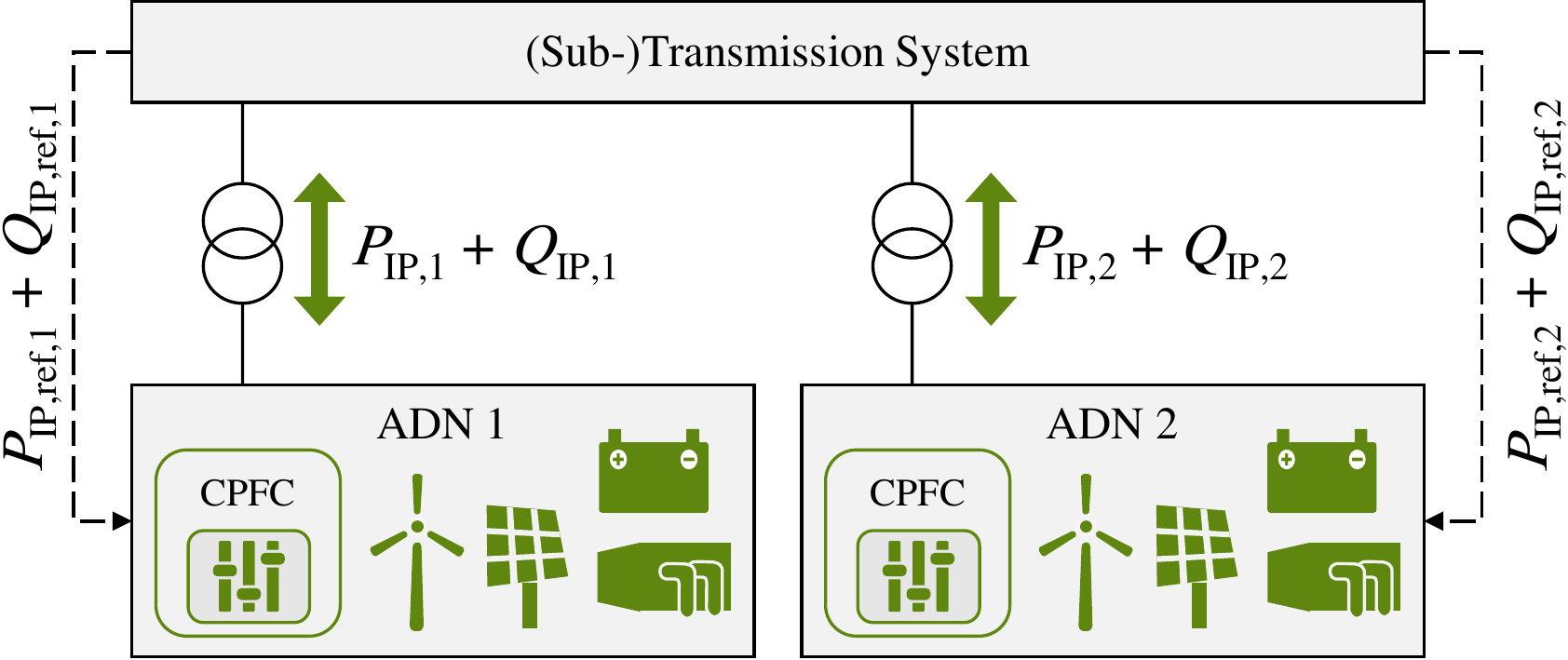}
	\caption{Active Distribution Network with Cross-Voltage-Level Power Flow Control in accordance with \cite{gonzalez2018flex}}
	\label{ADN}
\end{figure}

An ADN is a distribution grid whose network state can be adjusted by DERs. This property allows ADNs to provide ancillary services at the ADN's interconnection point to the next higher voltage level as illustrated in Fig. \ref{ADN}. The provision of ancillary services by ADNs has recently been the subject of numerous publications. Many approaches like \cite{stanojev2021primary,ValCustsem2013MPC,Pierre2017mang,parisio2017mpcmic} are based on optimal power flow calculations or model predictive control. In contrast, in \cite{zwartscholten2018hierarchical,gonzalez2017dynamic,robitzky2017impact,greve2015simulation}, PI-based ADN control concepts are presented. These ADNs are characterized by a CPFC that allows the real-time control of the ADN's power flow at the interconnection point to the next higher voltage level.

ADNs can be utilized for time-critical ancillary services in the emergency state where the tolerable response of the CPFC lies within seconds to minutes, such as:

\begin{itemize}
	\item Reactive / curative redispatch
	\item Frequency control (via provision of active power control by the underlying ADNs)
	\item Voltage control in transmission systems (via provision of reactive power control by the underlying ADNs)
	% \item Adjustment of the interconnection power flow in the context of emergency measures.
\end{itemize}

%g
The structure of the CPFC can be different. In \cite{gonzalez2017dynamic,robitzky2017impact}, the PI-Controller is implemented decentralized in each DER, whereas in \cite{zwartscholten2018hierarchical,greve2015simulation}, the controller is implemented centralized in the ADN. In an adaptive system, the central implementation has the advantage that only one controller needs to be adjusted. 

Because many DERs are connected to the Low Voltage (LV) or Medium Voltage (MV) grid, the coordination of the DERs is a complex challenge. As presented in \cite{zwartscholten2018hierarchical}, a hierarchical control structure can decrease the complexity of coordination. The High Voltage (HV) level ADN can rely on MV level ADNs for each subordinated distribution grid, whereby each ADN is controlled by its own control system. 

The fact that network restrictions could be violated by controlling DERs must be considered by designing the control system. 
In \cite{gonzalez2017dynamic}, the control system is blocked as long as voltage band violations occur until the on-load tap changer of the transformer at the interconnection point to the next higher voltage level returns the voltage to a permissible state. Besides, voltage band violations can occur, which the on-load tap changer can not remedy. In addition, line overloads in complex distribution networks are a challenge. In these cases, secondary control is necessary. For this purpose, an optimal power flow calculation, as described in \cite{Pierre2017mang,yuan2018lopf,anese2017acopf,dolan2012opftherm} could be used. Alternatively, the agent system from \cite{pohl2018sen,ulf2013phdagent} or protection and control devices from \cite{raj2020iauto} could be utilized for this task.

\section{Development and Structure of the Adaptive Control System}
\label{adapcontrol}

This section describes the development and structure of the adaptive control system. The actuators of the control loop are the DERs connected to the ADN. For the reasons mentioned in Section \ref{intro}, the availability of DERs changes over time. Theoretically, the set of actuators available to the control system can adopt any possible combination of DERs which participate in ADN's CPFC. Thus, the plant model of the ADN is a time-variant system, which can be described by a family of time-invariant systems. In order to meet this challenge, a gain scheduling approach is used.  Common use cases of gain scheduling are nonlinear systems like linear parameter-varying systems \cite{gahinet2013sched,apkarian1998sched}. In these approaches, the system’s operating range is divided into regions where linear control is adequate. For each region, a linear controller is designed. In \cite{jing2015sched,jang2008sched}, gain scheduling is also applied for time-variant systems, whereby the gain scheduling approach compensates the time-variance. Also in this work, the gain scheduling approach is used to compensate the time variance and not a nonlinear behavior.

For each possible set of actuators $A_n$, individual control parameters $K_\text{P}$ and $K_\text{I}$ are calculated in advance and saved in a lookup table $CP$. In order to choose the proper control parameters of $CP$, the control system has to know which DERs are available at all times. For this reason, the control system must regularly query the availability of the DERs.

Without any knowledge of the dynamic properties of the DERs that participate in the control loop, the plant model is a black box. The effort to create detailed models of the DERs for a white box approach is disproportionate. Therefore a grey box approach, which is a good trade-off between effort and accuracy, is chosen in this paper. After the structure of the adaptive control system is explained in Section \ref{structureCPFC}, a reference control loop model is created in Section \ref{refplant} by means of some data of the DERs. Finally, the reference control loop model is used to determine the parameters of the lookup table $CP$ in Section \ref{CP}.

\subsection{Structure of the Adaptive Control System}
\label{structureCPFC} 

\begin{figure}[t]
	\centering
	\includegraphics[scale=0.44]{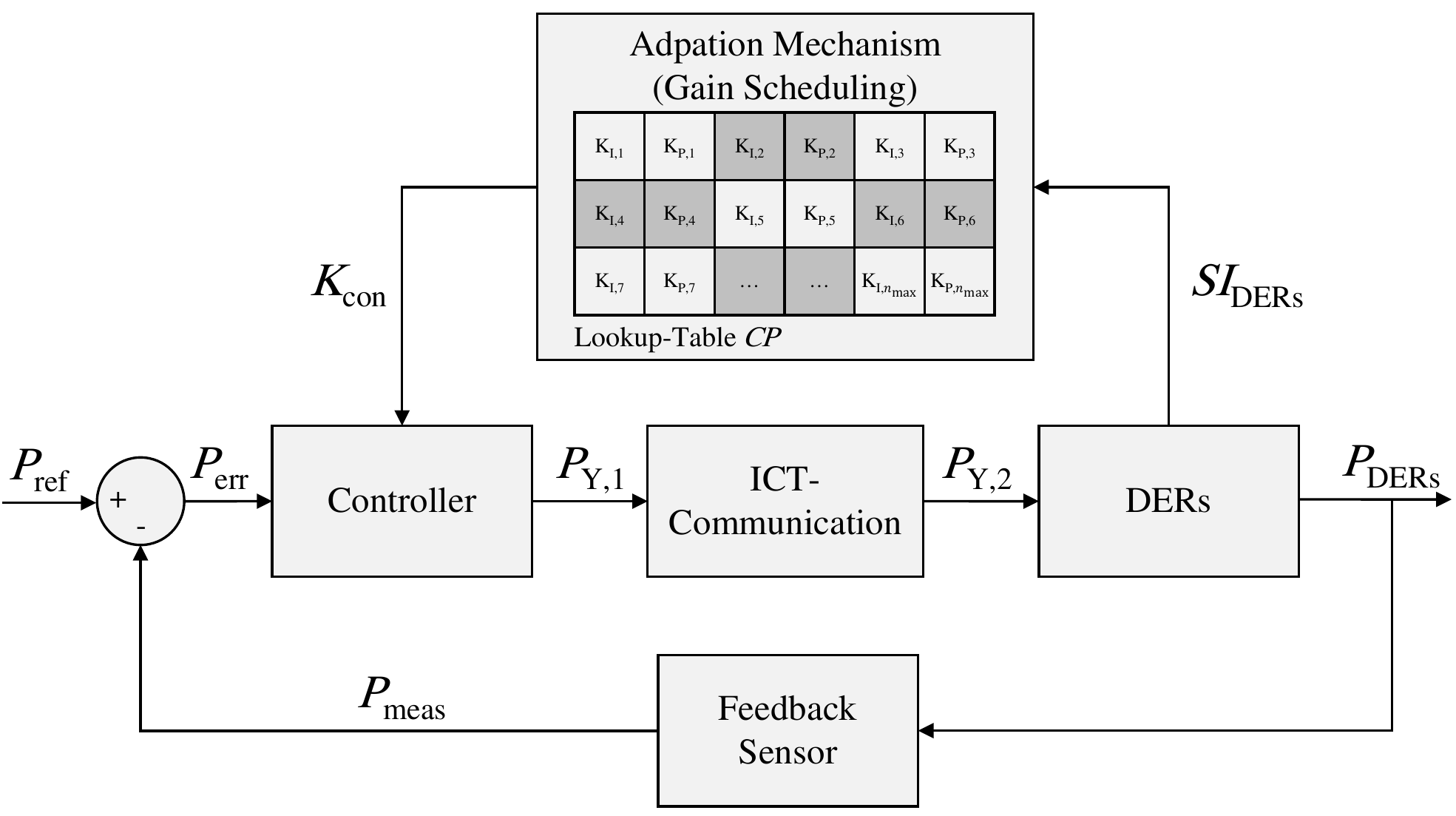}
	\caption{Control structure of the adaptive control system}
	\label{controlstucture}
\end{figure}

This paper extends the CPFC from \cite{zwartscholten2018hierarchical} by an adaption mechanism. Fig. \ref{controlstucture} illustrates an overview of the structure of the adaptive control system. A detailed description of the control system can be found in \cite{zwartscholten2018hierarchical}. In general, the control loop consists of a PI-based controller, ICT-communication, the DERs, and a feedback sensor. The controller gets the error signal $P_\text{err}$, which is the deviation between reference value $P_\text{ref}$ and measurement value $P_\text{meas}$. Based on $P_\text{err}$, the controller determines a correcting variable $P_\text{Y}$, which is transferred via ICT-communication to the DERs. Each DER has its own local controller that processes the signal $P_\text{Y}$. The setpoint of each DER $P_{\text{ref},d}^\text{DER}$ results from

\begin{equation}
	\begin{split}
		P_{\text{ref},d}^\text{DER}= P_\text{Y} \cdot C_d^\text{DER} + P_{\text{sched},d}^\text{DER}, 
	\end{split}
	\label{PDER}
\end{equation}
\noindent 

whereby $C_d^\text{DER}$ is the factor of participation in the control loop, and $P_{\text{sched},d}^\text{DER}$ is the schedule of the DER $d \in D$, where $D$ is the set of all DERs that participate in the control loop. In this work, $C_d^\text{DER}$ is set to the installed power of each DER. Based on $P_\text{Y}$, the DERs change their infeed into the electrical grid. The goal of the control system is to control the interconnection power flow $P_\text{IP}$ to a higher voltage level. The electrical grid and the loads are not considered in the structure of Fig. \ref{controlstucture}. This simplifications are justified in Section \ref{refplant}. Due to the simplifications, the control loop shown in Fig. \ref{controlstucture} controls the infeed of all DERs $P_\text{DERs}$ instead of $P_\text{IP}$.          

The lookup table $CP$ of the gain scheduling based adaption mechanism contains one pair of the control parameters $K_{\text{I},n}$ and $K_{\text{P},n}$ (summarized as $K_{\text{con}}$) for each possible set of actuators $A_{n}$. In order to choose the right parameter pair, the adaption mechanism has to know which DERs are available to the control system. The status information $SI_\text{DERs}$ provides these data. The signal $SI_\text{DERs}$ contains a boolean for each DER $d$. The element $d$ of $SI_\text{DERs}$ is zero if the DER $d$ is unavailable and one if the DER $d$ is available to the control system.

\begin{figure}[htp]
	\centering
	\includegraphics[scale=0.44]{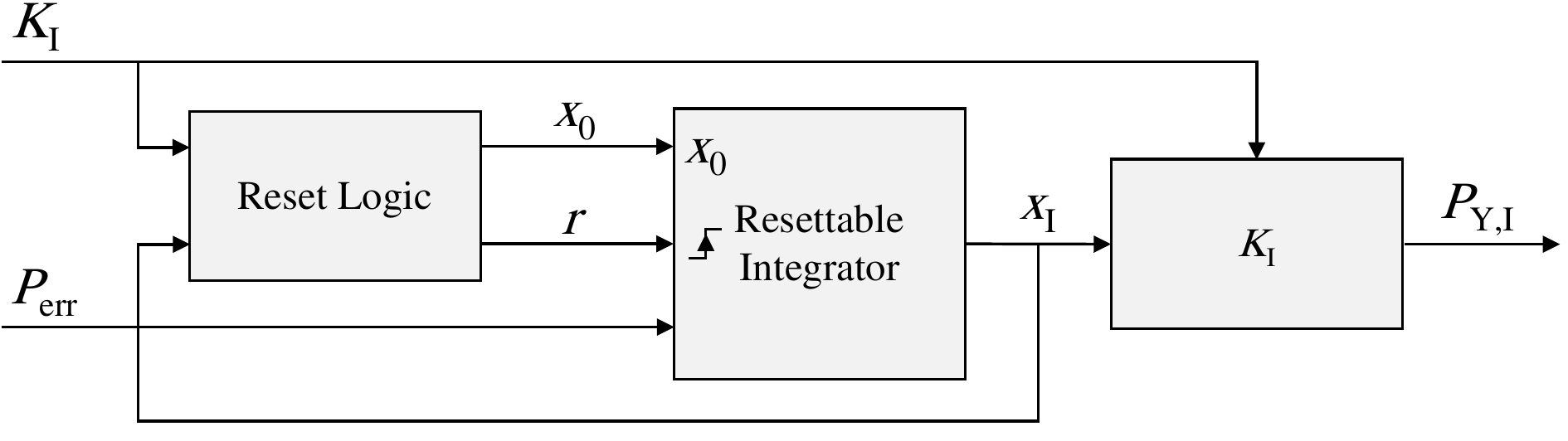}
	\caption{Control structure extension to avoid control errors after control parameter switching}
	\label{stucturecpswitch}
\end{figure}

In order to avoid unwanted control errors by switching the control parameters, a reset logic is added to the control structure of the integral term.
The reset logic, shown in Fig. \ref{stucturecpswitch}, realizes a reset of the integrator so that the output value of the integral term $P_\text{Y,I}$ remains unchanged when switching the control parameters. For this purpose, the reset logic uses the state variable of the integrator $x_\text{I}$ and the specified integral control parameter of the adaption mechanism $K_\text{I}$ as input parameters. 

The output parameter $x_\text{0}$ of the reset logic in discrete-time representation is determined by Eq. \ref{KIincrease} to Eq. \ref{resetlogic}.

\begin{equation}
	x_\text{0,inc}^{(k)}= x_\text{I}^{(k-1)}+ x_\text{I}^{(k-1)} \cdot (K_\text{I}^{(k-1)}-K_\text{I}^{(k)}) \cdot K_\text{I}^{(k)}
	\label{KIincrease}
\end{equation}

\begin{equation}
	x_\text{0,dec}^{(k)}= 	x_\text{I}^{(k-1)}+ x_\text{I}^{(k-1)} \cdot \frac{K_\text{I}^{(k-1)}-K_\text{I}^{(k)}}{K_\text{I}^{(k)}}
	\label{KIdecrease}
\end{equation}

\begin{equation}
	x_\text{0}^{(k)}=
	\begin{cases}
		x_\text{0,inc}^{(k)} & \text{if }K_\text{I}^{(k)} > K_\text{I}^{(k-1)}\\
		x_\text{0,dec}^{(k)} & \text{if }K_\text{I}^{(k)} \leq K_\text{I}^{(k-1)}.
	\end{cases}
	\label{resetlogic}
\end{equation}
\noindent
The parameter $x_\text{I}^{(k-1)}$ is the state variable of the integrator and $K_\text{I}^{(k-1)}$ is the gain of the integral term immediately before the reset. The rising edge of the reset signal $r$, described by 

\begin{equation}
	r=
	\begin{cases}
		0   & \text{if } K_\text{I}^{(k)} = K_\text{I}^{(k-1)}\\
		1       & \text{if } K_\text{I} \neq K_\text{I}^{(k-1)},
	\end{cases}
	\label{resetsignal}
\end{equation}
\noindent
triggers the integrator to apply $x_{0}^{(k)}$ as the new state variable $x_\text{I}^{(k)}$. 

\subsection{Reference Control Loop Model}
\label{refplant}

For the calculation of the lookup table $CP$ and the analysis in Section \ref{results}, reference control loop models are used. In order to implement the reference control loop model, the following parameters of the DERs are necessary:

\begin{itemize} 
	\item installed active power $P_{\text{inst},d}^\text{DER}$ (for Q-control $Q_{\text{inst},d}^\text{DER}$), 
	\item reaction time $T_{\text{delay},d}^\text{DER}$, and
	\item rise time $T_{\text{rise},d}^\text{DER}$ (from 10\,\% to 90\,\%).
\end{itemize} 
\noindent   
Each DER $d$ is represented by the transfer function

\begin{equation}
	G_{\text{PT1},d}^\text{DER}(s) = e^{-T_{\text{delay,}d}^\text{DER} \cdot s} \cdot \frac{K_{d}}{1+T_{d}^\text{DER} \cdot s}. 
	\label{GDER}
\end{equation}
\noindent
Eq. \ref{GDER} consists of a delay and a first-order transfer function. Based on the rise time $T_{\text{rise},d}^\text{DER}$, an equivalent time constant $T_{d}^\text{DER}$ for the first-order transfer function can be calculated by

\begin{equation}
	T_{d}^\text{DER} = \frac{T_{\text{rise},d}^\text{DER}}{2.197}. 
	\label{TDER}
\end{equation}
\noindent

The denominator of 2.197 can be determined from the step response of any first-order transfer function. The time constant $T_{d}^\text{DER}$ is calculated using $T_{\text{rise},d}^\text{DER}$, since the rise time can be determined directly from a step response. Thus, an equivalent first-order transfer function can be determined from a step response (see Section \ref{modelling}). The gain $K_d$ is described by

\begin{equation}
	K_{d} = w_a \cdot P_{\text{inst},d}^\text{DER}, 
	\label{Kd}
\end{equation}
\noindent
where $w_a$ ($a \in A_n$) is an individual scaling factor of each DER. For each possible set of actuators $A_n$, a reference plant model is created by 

\begin{equation}
	G_{A_n}^\text{DERs}(s) = \sum_{a \in A_n} G_{a}^\text{DER}(s). 
	\label{GADER}
\end{equation}
\noindent
The open-loop transfer function is

\begin{equation}
	G_{A_n}^\text{OL}(s) = G_\text{PI}(s) \cdot G_\text{Com}(s) \cdot 	G_{A_n}^\text{DER}(s).
	\label{GOL}
\end{equation}
\noindent
The PI-Controller is described by 

\begin{equation}
	G_\text{PI}(s) = \frac{\frac{K_\text{P}}{K_\text{I}} \cdot s + K_\text{P}} {\frac{K_\text{P}}{K_\text{I}} \cdot s},
	\label{GPI}
\end{equation}
\noindent
the ICT-communication by

\begin{equation}
	G_\text{Com}(s) = e^{-T_\text{com} \cdot s}, 
	\label{GCOM}
\end{equation}
\noindent
and the feedback sensor by 

\begin{equation}
	G_\text{M}(s) = e^{-T_\text{meas} \cdot s},
	\label{GM}
\end{equation}
\noindent
where $T_\text{com}$ is the communication delay and $T_\text{meas}$ is the measurement delay. The closed-loop transfer function of the reference control loop model for each set $A_n$ is expressed by

\begin{equation}
	G_{A_n}^\text{CL}(s) = \frac{G_{A_n}^\text{OL}(s)}{1+G_{A_n}^\text{OL}(s) \cdot G_\text{M}(s)}.
	\label{GCL}
\end{equation}
\noindent
The number of possible sets of DERs $n_\text{max}$ results from

\begin{equation}
	n_\text{max} = 2^m - 1,
	\label{nmax}
\end{equation}
\noindent
where $m$ is the total number of DERs participating in the control loop. The electrical grid and all loads are not considered in the reference control loop model. In a congestion-free grid, the electrical lines affect the control loop through the line losses and their reactive power behavior. The loads affect the control loop via the voltage dependency. As shown in \cite{zwartscholten2020impact}, the electrical grid and the load have only a minor impact on the control system's performance and stability. Therefore, the electrical grid and all loads are neglected in the reference control loop model, in order to reduce computational costs.

\subsection{Design of the Lookup Table}
\label{CP}
The reference control loop models are used to determine the elements of the lookup table $CP$. In doing so, the disk margin and overshoot are considered. Compared to the gain and phase margin, the disk margin is the stronger design criterion for robust control \cite{matlab2021robust}. In order to create a highly robust system, the control parameters in $CP$ are tuned by the necessary condition disk margin $\geq$\,1.5. To get a smooth performance of the control loop, also overshooting $<$\,1$\,\%$ is considered as a necessary condition. 

\begin{figure}[htp]
	\centering
	\includegraphics[scale=0.53]{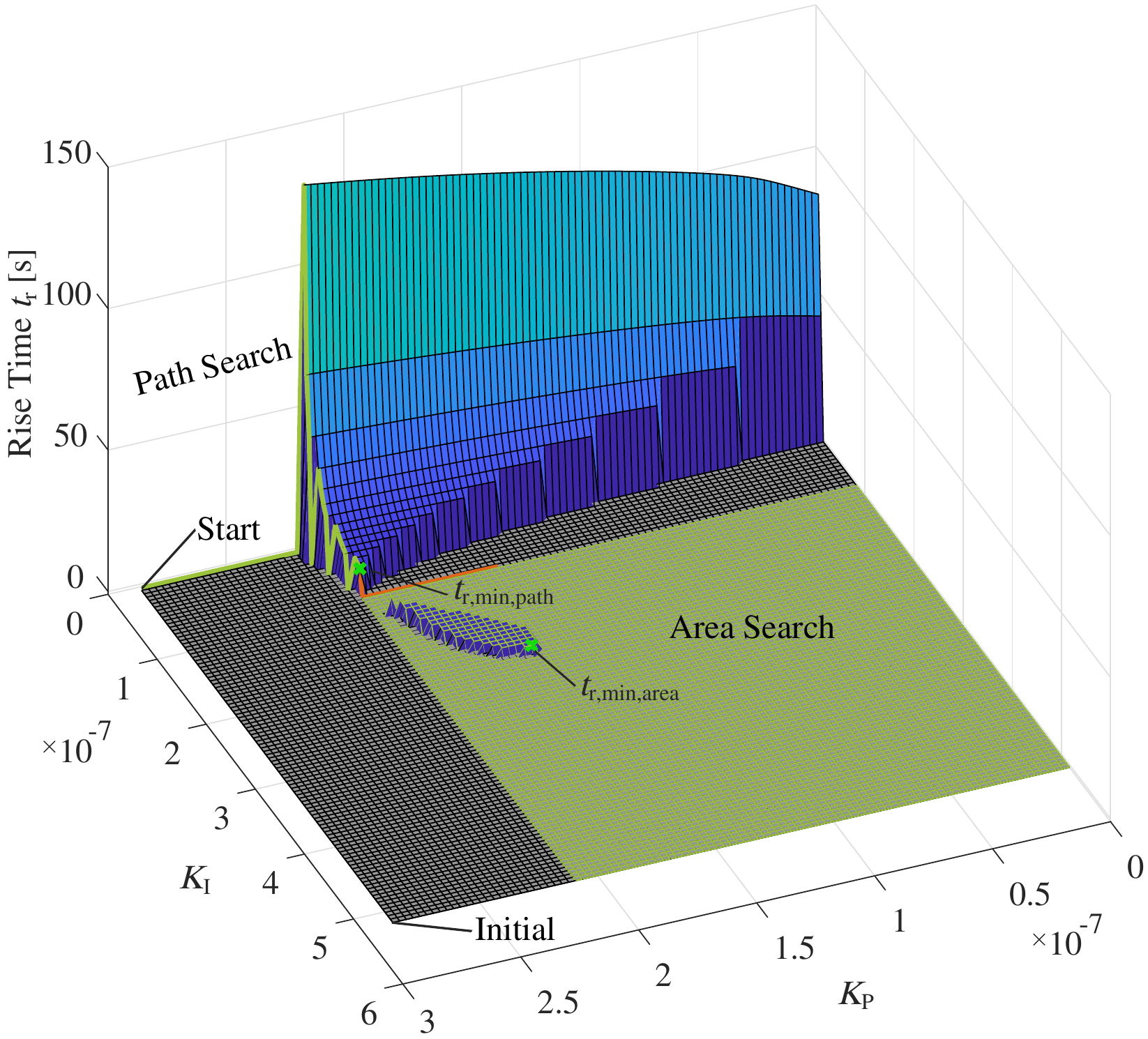}
	\caption{Exemplary non-convex solution space}
	\label{solutionspace}
\end{figure}

The feasible solution space (disk margin $\geq$\,1.5 and overshooting $<$\,1$\,\%$) for $K_\text{P}$ and $K_\text{I}$ of an exemplary reference control loop model is sketched in Fig. \ref{solutionspace}. The dark grey plane marks the non-feasible solutions. For the feasible solutions, the rise time $t_\text{r}$ is given on the z-axis. The goal is to find the feasible solution with the lowest rise time of every combination of DERs. For this purpose, an algorithm has been developed, which is explained in the following.  

\begin{figure}[h]
	\centering
	\includegraphics[scale=0.5]{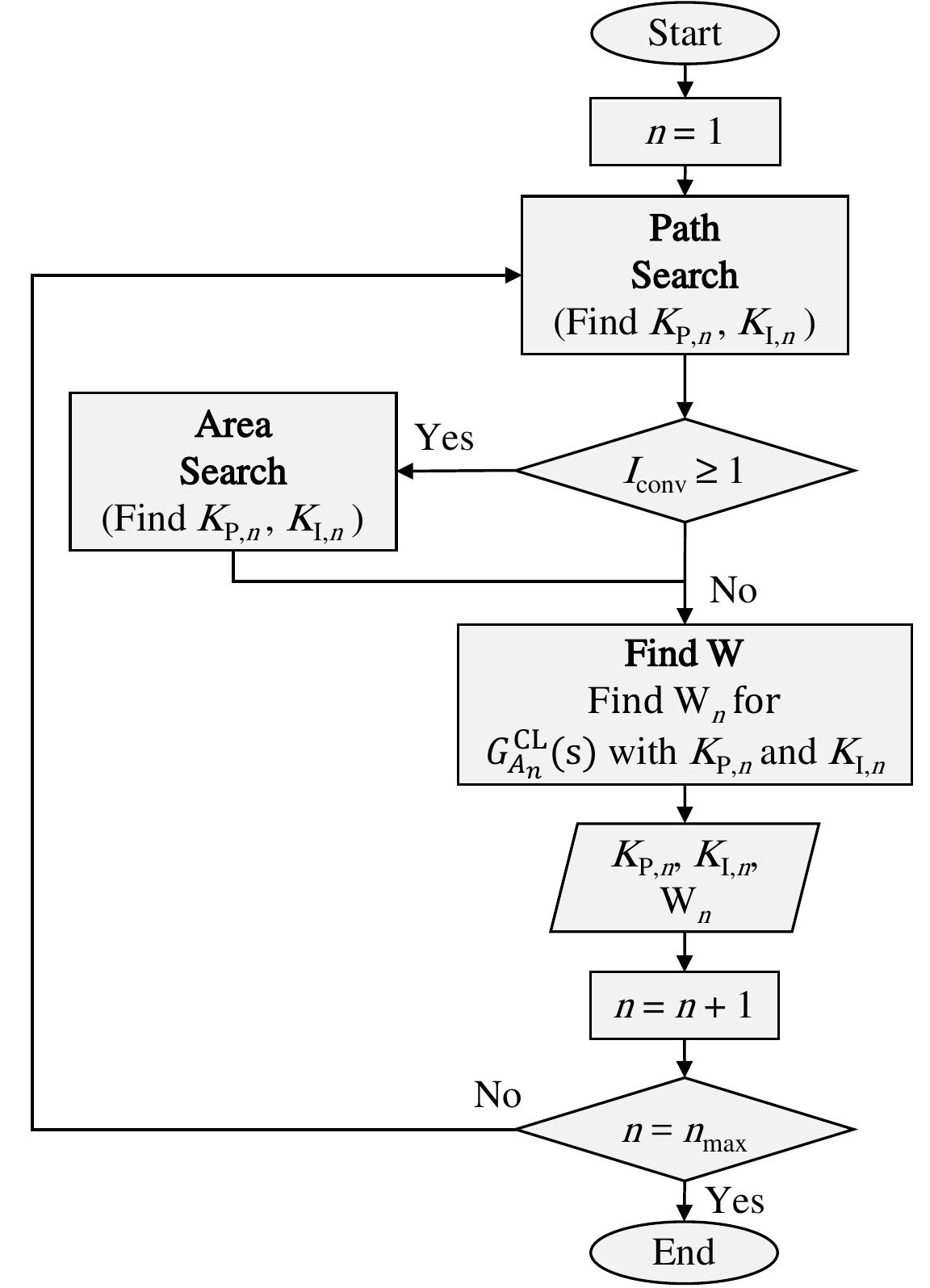}
	\caption{Flow chart of the main function for determining the lookup table $CP$ and the sets of scaling factors $W_n$}
	\label{flowchart}
\end{figure}

The inputs of the algorithm are $P_{\text{inst},d}^\text{DER}$, $T_{\text{delay},d}^\text{DER}$, and $T_{\text{rise},d}^\text{DER}$ for each DER (compare Table \ref{DERpara}) and the outputs are $CP$ and a set of individual scaling factors $W_n$. The array $W_n$ contains the scaling factors $w_a$ for each DER $a$ of a combination $A_n$. In the beginning, all arrays $W_n$ contain only ones. The flowchart of the main function of the algorithm is shown in Fig. \ref{flowchart}.

\begin{figure*}[h]
	\centering
	\includegraphics[scale=0.35]{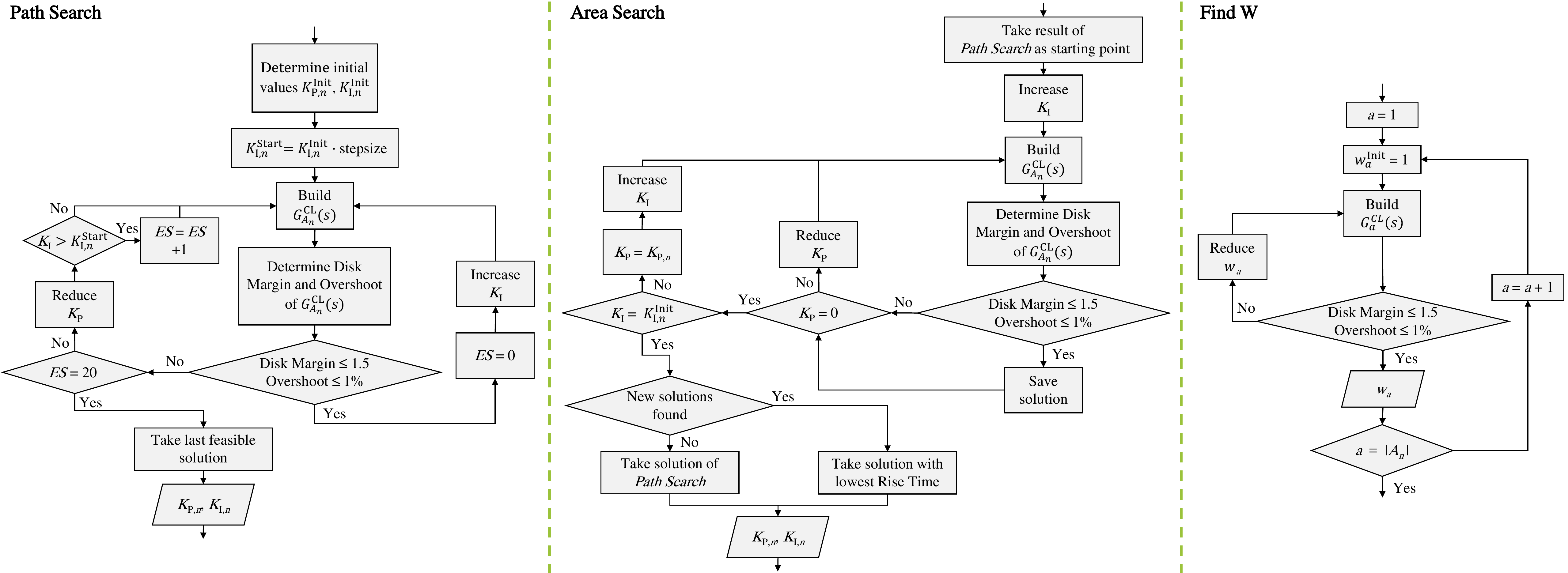}
	\caption{Flow chart of the sub-functions for determining the lookup table $CP$ and the sets of scaling factors $W_n$}
	\label{flowchart2}
\end{figure*}

The algorithm successively searches the parameters $K_\text{P}$, $K_\text{I}$, and $w_a$ starting with the combination $A_1$ ($n = 1$). The main function contains the three sub-functions \textit{Path Search}, \textit{Area Search}, and \textit{Find W}. Fig. \ref{flowchart2} illustrates these three sub-functions. In addition, the procedures of the sub-functions \textit{Path Search} and \textit{Area Search} are marked in Fig. \ref{solutionspace} in green. The sub-function \textit{Path Search} is called first. In the beginning, the sub-function \textit{Path Search} determines initial values of $K_\text{P}$ and $K_\text{I}$. The initial values must be outside of the feasible solution space. In this work, the Matlab function \textit{pidtune} with default options is used for this purpose. The initial values determined by \textit{pidtune} do not meet the high robustness requirements for the control system regarded here, which is why these parameters are suitable as initial parameters.

In the next step, $K_\text{I}^\text{Start}$ is set to the product of $K_\text{I}^\text{Init}$ and the step size. It should be mentioned that the algorithm is performed successively with two different step sizes. In a first run, the step size is 0.01, and in a second run, the step size is 0.001. In the second run, the initial and start values of $K_\text{P}$ and $K_\text{I}$ are set to the output of the first run. In this way, computing time can be reduced. After setting $K_\text{I}^\text{Start}$, \textit{Path Search} starts from the upper left corner of the examined solution space (see Fig. \ref{solutionspace}).

Next, \textit{Path Search} builds $G_{A_n}^\text{CL}(s)$ according to Eq. \ref{GDER} to Eq. \ref{GCL} ($\forall w_a = 1$) and determines the disk margin and overshooting. The disk margin and overshooting are calculated with the Matlab functions \textit{diskmargin} and \textit{stepinfo} \cite{matlab2021robust}. Then $K_\text{P}$ is reduced according to Eq. \ref{redu} until a feasible solution is found. In doing so, the algorithm checks whether a feasible solution has already been found by comparing $K_\text{I}$ with $K_\text{I}^\text{Start}$ ($K_\text{I}$ $>$ $K_\text{I}^\text{Start}$). As soon as a feasible solution is found, the parameter $K_\text{I}$ is increased according to Eq. \ref{incr} until again a non-feasible solution is found.

In the following, the algorithm reduces $K_\text{P}$ if a non-feasible solution is found and increases $K_\text{I}$ if a feasible solution is found.
By this procedure, the function searches along a path. The path always leads along the left edge of the feasible solution space. When $K_\text{P}$ has been increased 20 times in a row, which is counted by the parameter $ES$, the search is finished (see orange path in Fig. \ref{solutionspace}). 
Empirical investigations have shown that the global minimum is always located at the lower-left tip of the feasible solution space as long as the feasible solution space is convex. Therefore the last feasible solution found by \textit{Path Search} is the global minimum in a convex solution space. In rare cases, the feasible solution spaces are non-convex, as shown in Fig. \ref{solutionspace}. In these cases, the function \textit{Path Search} can not find the global minimum. For this reason, the main function checks after running \textit{Path Search} whether the solution space may be non-convex. 

This is done with the indicator $I_\text{conv}$, which is calculated by Eq. \ref{Timecona} and Eq. \ref{conv}, whereby the rise time $t_\text{r,min,path}$ of the solution of the \textit{Path Search} function is used. If $I_\text{conv}$ $\geq$ 1 is true, the main function calls the function \textit{Area Search}, which needs significantly more computing power compared to \textit{Path Search}. Empirical evaluations have shown that for non-convex solution spaces, $I_\text{conv}$ is always higher than one.

\begin{equation}
	K_\text{P} = K_\text{P} - K_\text{P} \cdot \text{stepsize},
	\label{redu}
\end{equation}
\noindent

\begin{equation}
	K_\text{I} = K_\text{I} + K_\text{I} \cdot \text{stepsize},
	\label{incr}
\end{equation}
\noindent

\begin{equation}
	T_\text{avg} = \frac{\sum_{a \in A_n} ((T_{\text{delay},a}^\text{DER} + T_{\text{rise},a}^\text{DER}) \cdot P_{\text{inst},a}^\text{DER})}{\sum_{a \in A_n} P_{\text{inst},a}^\text{DER}},
	\label{Timecona}
\end{equation}
\noindent

\begin{equation}
	I_\text{conv} = \frac{t_\text{r,min,path}}{T_\text{avg} + T_\text{meas} + T_\text{com}}.
	\label{conv}
\end{equation}
\noindent

The function \textit{Area Search} starts with the solution found by \textit{Path Search} and increases $K_\text{I}$ according to Eq. \ref{incr} in the beginning. Subsequently, the function searches the whole rectangle until $K_\text{P}$ reaches zero and $K_\text{I}$ reaches $K_\text{I}^\text{Init}$. If \textit{Area Search} finds new solutions, the solutions are saved. After the search is finished, the solution with the lowest rise time $t_\text{r}$ is taken as the output of \textit{Area Search}. If \textit{Area Search} does not find any new solution, the output of \textit{Area Search} is the solution of \textit{Path Search}.  

After $K_\text{I}$ and $K_\text{P}$ are determined for the combination $n$, the main function calls the sub-function \textit{Find W} to find a set of scaling factors $W_n$ for each combination $n$. For this purpose, the closed-loop transfer function $G_a^\text{CL}(s)$ is determined for each DER $a \in A_n$ according to Eq. \ref{GDER} to Eq. \ref{GCL}, whereby Eq. \ref{GADER} can be ignored. In Eq. \ref{GPI}, the previously determined controller parameters $K_{\text{P},n}$ and $K_{\text{I},n}$ are used. Afterward, the function checks if $G_a^\text{CL}(s)$ fulfills the disk margin and overshooting requirements. If the requirements are not met, $w_a$ is reduced stepwise by 0.1 \% of  $w_{a}^\text{Init}$. The individual scaling factors $w_a$ reduce the influence of DERs, which are not conducive to the robustness of the control loop. In this way, the robustness of the control loop is further increased.

\section{Overview of the Simulations Models}
\label{modelling}
\begin{figure}[h]
	\centering
	\includegraphics[scale=0.35]{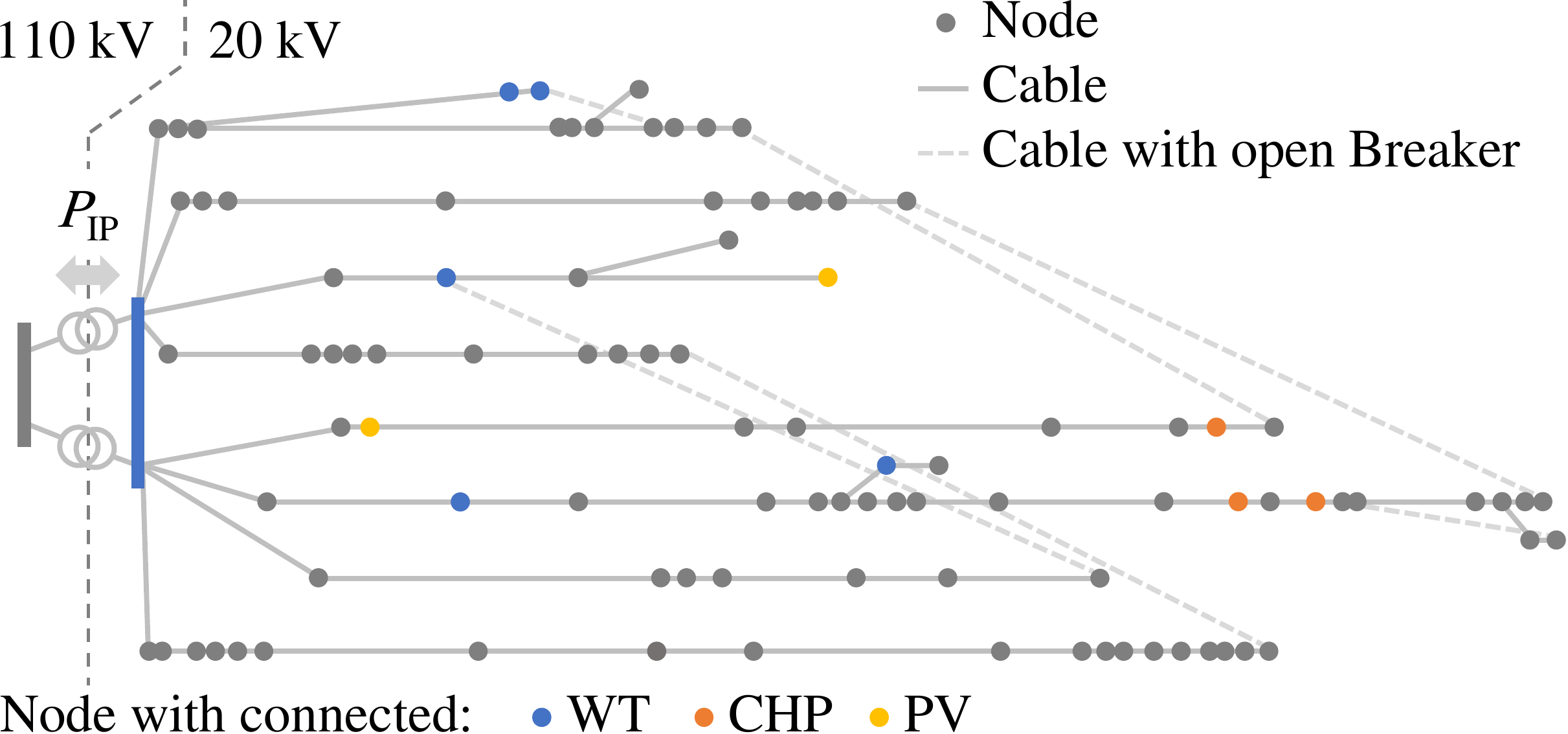}
	\caption{SimBench 20\,kV rural benchmark grid \cite{simbench2021grid}}
	\label{simbench}
\end{figure}

In order to analyze the adaptive control system, the SimBench 20\,kV rural benchmark grid (SimBench Code: 1-MV-rural-0-no\_sw) is used \cite{simbench2021grid}. Eleven DERs, which are directly connected to the 20\,kV grid, are integrated as actuators of the control system. The connection points of these DERs are colored in the overview of the Simbench grid in Fig. \ref{simbench}. In the SimBench rural benchmark grid, Wind Turbines (WTs), Combined Heat and Power (CHP) plants, and photovoltaic (PV) plants are connected.
In this work, two different modeling methods are applied. For the determination of $CP$, $W_n$, and the analysis in Section \ref{robper} and \ref{Analy_Uncertain}, the reference control loop model is used. In order to validate the determination of $CP$ and $W_n$ based on the LTI modeling, simulations are performed using a detailed simulation model of the Simbench grid in Matlab Simulink (see Section \ref{timdom}).

 ~\linebreak
\noindent
\textbf{Reference Simulation Model}

\begin{figure}[h]
	\centering
	\includegraphics[scale=0.48]{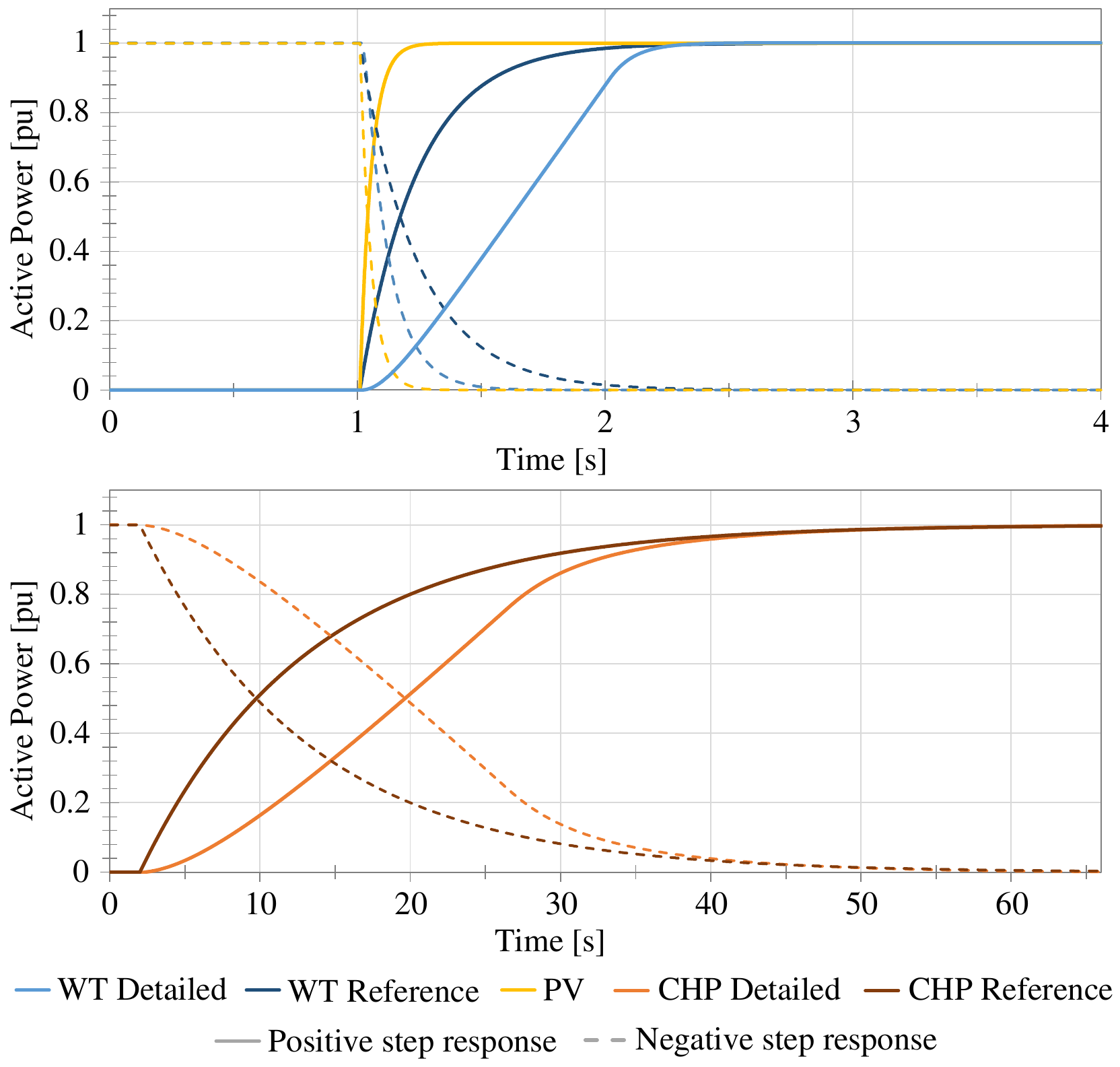}
	\caption{Step response of the detailed and reference DER models}
	\label{SRMod}
\end{figure}

The reference simulation model is based on the LTI modeling from Section \ref{refplant}. The input parameters of the DERs are shown in Table \ref{DERpara}. The parameters $P_{\text{inst},d}^\text{DER}$ are given in the Simbench dataset, whereas $T_{\text{delay},d}^\text{DER}$ and $T_{\text{rise},d}^\text{DER}$ were assumed by using the DER models of the detailed simulation model. For each DER-type, $T_{\text{rise},d}^\text{DER}$ is determined by the medium rise time of the positive and negative step response of the detailed model of the corresponding DER-type. In order to demonstrate the dynamic behavior of the reference DER models and the detailed DER models, the step responses (at $t$ = 1) are contrasted in Fig. \ref{SRMod}. The detailed CHP plant model is characterized by comparatively large time constants $T_{\text{delay},d}^\text{DER}$ and $T_{\text{rise},d}^\text{DER}$. The peculiarity of the detailed WT model is that the negative and positive step responses have a different rise time. In addition, a clearly recognizable deviation from the response of the first-order transfer function can be observed in the step responses of the detailed CHP plant and WT model. The ICT-communication and the feedback sensor delays are assumed with $T_\text{com}$ = $T_\text{meas}$ = 0.1\,s.  

\begin{table}[htbp]
	\caption{Input parameters of the DERs}
	\begin{center}
		\begin{tabular}{|c|c|c|c|c|}
			\hline
			\textbf{DER-}& \textbf{DER-} &\multicolumn{3}{|c|}{\textbf{Input Parameters}} \\
			\cline{3-5} 
			\rule{0pt}{8pt}\textbf{Name} & \textbf{Type} & \textbf{\textit{P}$_{\text{inst}\textbf{,\textit{ d}}}^\text{DER}$ [W]}& \textbf{\textit{T}$_{\text{delay}\textbf{,\textit{ d}}}^\text{DER}$ [s]}& \textbf{\textit{T}$_{\text{rise}\textbf{,\textit{ d}}}^\text{DER}$ [s]} \\
			\hline
			SGen 1 & WT & 2E6 & 0.01 & 0.515  \\
			\hline
			SGen 2 & WT & 2E6 & 0.01 & 0.515  \\
			\hline
			SGen 3 & WT & 1.7E6 & 0.01 & 0.515  \\
			\hline
			SGen 4 & WT & 1.9E6 & 0.01 & 0.515 \\
			\hline
			SGen 5 & CHP & 0.31E6 & 1 & 24.57 \\
			\hline
			SGen 6 & CHP & 0.35E6 & 1 & 24.57 \\
			\hline
			SGen 7 & WT & 1.8E6 & 0.01 & 0.515 \\
			\hline
			SGen 8 & WT & 2E6 & 0.01 & 0.515 \\
			\hline
			SGen 9 & PV & 0.195E6 & 0.01 & 0.1  \\
			\hline
			SGen 10 & PV & 0.125E6 & 0.01 & 0.1 \\
			\hline
			SGen 11 & CHP & 0.28E6 & 1 & 24.57 \\
			\hline
		\end{tabular}
		\label{DERpara}
	\end{center}
\end{table} 
\noindent
\textbf{Detailed Simulation Model} 

In this model, the CHPs are modeled according to \cite{strunck2021chp} and the WTs according to IEC 61400-27-1 (Type 4B). The PV systems' setpoint tracking are assumed to behave like a power inverter. In \cite{greve2015simulation,gonzalez2017dynamic}, the dynamic behavior of inverter-based DERs is modeled as a first-order transfer function. Furthermore, the first-order response of an inverter is demonstrated by numeric simulations in \cite{lu2019inve}. For this reason, the PV plants used in this work are modeled as first-order transfer functions, including a small delay for signal processing. Therefore, the dynamic behavior of the PV plants is modeled similarly to the reference model in Section \ref{refplant} plus a current source as an interface to the electrical grid. All other generation units that are not integrated as control system actuators are implemented as constant current sources. The upstream 110\,kV grid is implemented as an ideal voltage source. Furthermore, loads are modeled as constant impedances. For lines, the three-phase section line model from Simscape Power Systems is used. The simulations are executed with the phasor simulation method and Heun’s integration technique. 

\section{Analysis and Simulation Results}
\label{results}
This section is structured as follows. First, the robustness and performance of the adaptive control system are analyzed. In order to illustrate the advantages of adaptive control, a comparison is made with a static control system. Afterward, the influence of uncertainties is evaluated. All these investigations are performed using the reference simulation model. Finally, the adaptive control system is demonstrated by means of time-domain simulations using the detailed simulation model.

\subsection{Robustness and Performance Analysis}
\label{robper}
  
\begin{figure}[b]
	\centering
	\includegraphics[scale=0.52]{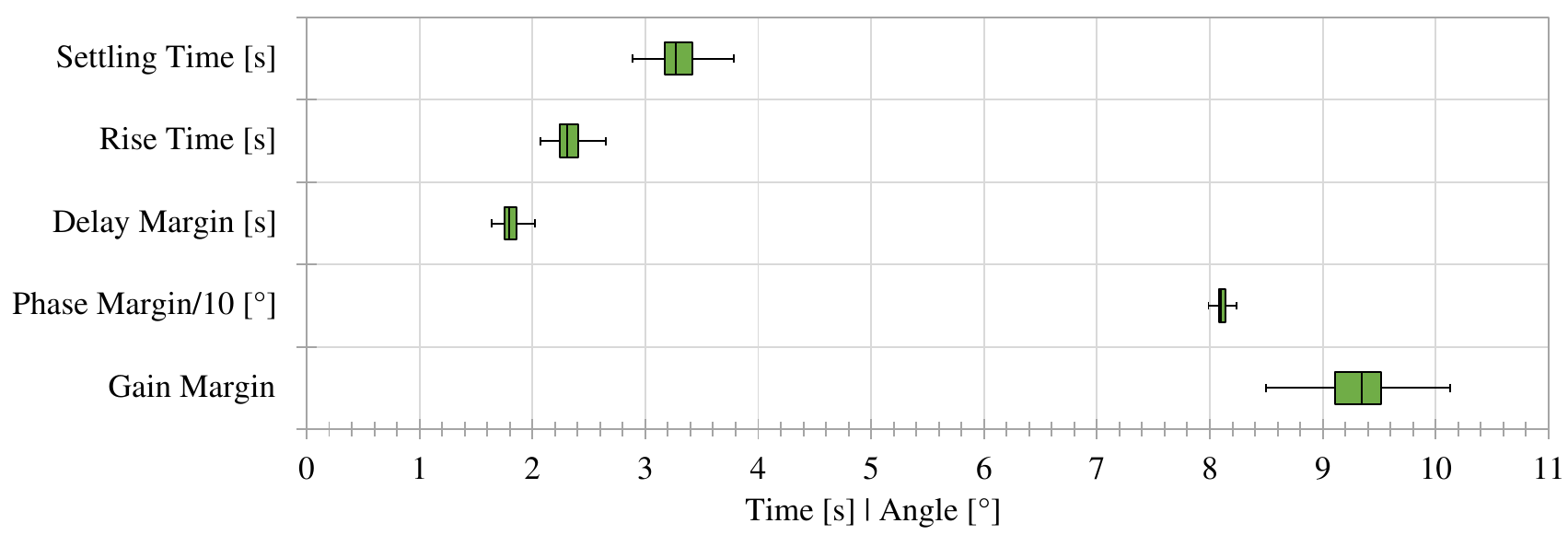}
	\caption{Boxplot of robustness and performance parameters of the adaptive controller}
	\label{Boxadapt}
\end{figure}

In this subsection, the robustness and performance of the adaptive control system are analyzed using the reference model described in Section \ref{refplant}. For this purpose, for all possible DER combinations $A_n$ ($n_\text{max} = 2047$), robustness and performance parameters are calculated by means of the Matlab functions \textit{allmargin}, \textit{discmargin}, and \textit{stepinfo} \cite{matlab2021robust}. The results of the adaptive control system are plotted in Fig. \ref{Boxadapt} as Boxplots, whereby outliers are neglected. The settling time, with a tolerance of 5\%, and the rise time are in the single-digit seconds range, which demonstrates the control system's fast response. Furthermore, the robustness parameters show a high reserve of delay, phase, and gain until stability limits are reached.

\begin{figure}[b]
	\centering
	\includegraphics[scale=0.52]{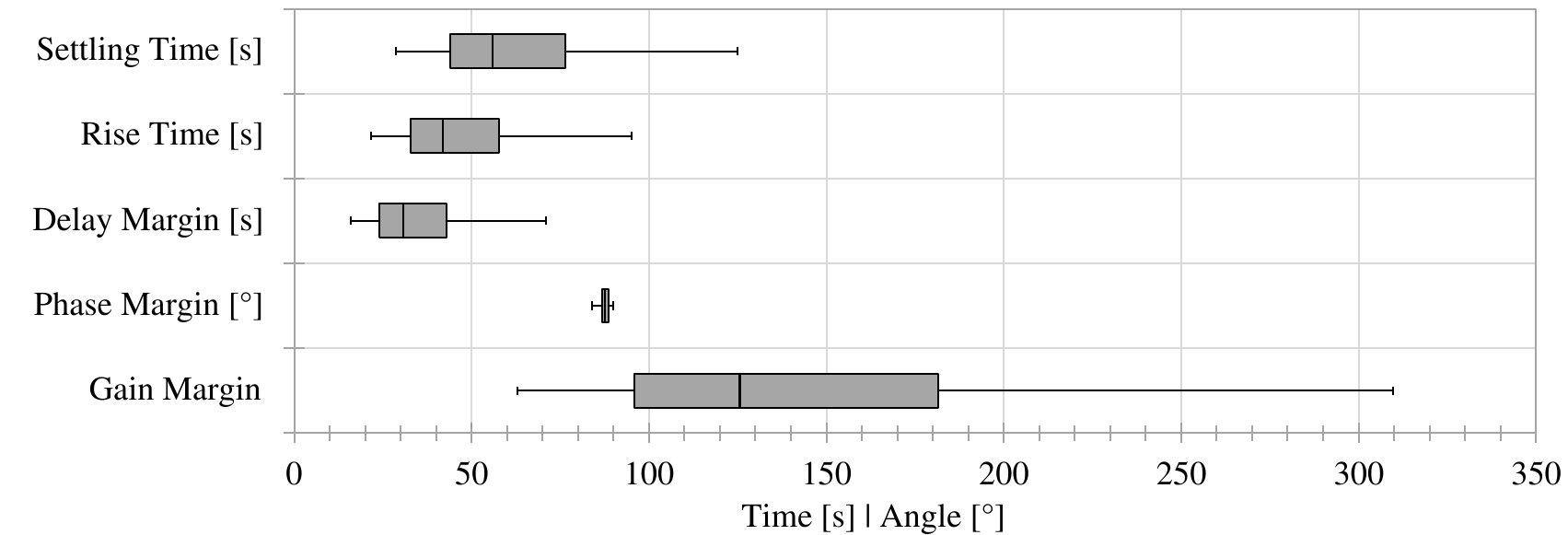}
	\caption{Boxplot of robustness and performance parameters of the static controller}
	\label{Boxstat}
\end{figure}

In Fig. \ref{Boxstat}, the same Boxplots are illustrated for the static controller. The static controller uses one pair of control parameters for all combinations of actuators. The static control parameters are chosen to meet the same requirements as the adaptive controller (disk margin $\geq$\,1.5 and overshooting $<$\,1$\,\%$) for all combinations $A_n$. The comparison of the performance parameters rise time and settling time illustrates the potential of the adaptive control system. Through the conservative setting of the control parameters ($K_\text{P} = 5.38\text{E}-9$, $K_\text{I} = 8.18\text{E}-9$), the static controller's settling time and rise time are several times higher. At this point, it should be noted that the control parameters seem to be very small. The reason is that the control loop is designed in SI units, so that the signal $P_\text{Y}$ is multiplied by the installed power $P_{\text{inst},d}^\text{DER}$ in the units watt (see Eq. \ref{Kd}). The static controller's conservative design increase all three margins. The margins are unnecessarily high at the expense of the performance. 

\subsection{Analysis of Model Uncertainties}
\label{Analy_Uncertain}

In addition, robustness is analyzed by considering model uncertainties. The influence of additional gain and phase has already been considered in Section \ref{robper}. This subsection focuses on the influence of deviations of the time constant $T_{\text{unc},d}^\text{DER}$ and deviations from a first-order behavior of the DER models. In order to analyze the deviations of the time constants, uncertain time constants $T_{\text{unc},d}^\text{DER}$ are used. The time constant $T_{\text{unc},d}^\text{DER}$ is described by

\begin{equation}
	T_{\text{unc},d}^\text{DER} = T_{d}^\text{DER} \cdot b, 
	\label{Tunc}
\end{equation}
\noindent
whereby $b$ is the factor of uncertainty. In this section, the parameters of $CP$ are still determined by considering $T_{\textit{ d}}^\text{DER}$ in the reference model of Section \ref{refplant}. But now, in the analyses of the control loop, the uncertain time constants $T_{\text{unc},d}^\text{DER}$ are used in Eq. \ref{GDER} to Eq. \ref{GCL}. The result is a closed-loop transfer function of the reference plant model $G_{A_n,\text{unc}}^\text{CL}(s)$ with uncertainties. Thus, the influence of deviations between the parameters used for the design of the control system and the uncertainty parameters are evaluated. In further investigation, Eq. \ref{GDER} is replaced by Eq. \ref{GPT2} or Eq. \ref{GRHPZ}. First, all actuators are assumed to behave like a second-order transfer function (model type PT2), described by   

\begin{equation}
	G_{\text{PT2},d}^\text{DER}(s) = e^{-T_{\text{delay,}d}^\text{DER} \cdot s} \cdot \frac{K_{d}}{(\frac{T_{\text{unc},d}^\text{DER}}{1.5291})^2 \cdot s^2 + \frac{T_{\text{unc},d}^\text{DER}}{0.7645} \cdot s + 1}. 
	\label{GPT2}
\end{equation}
\noindent
Subsequently, all actuators are modeled as Right-Half-Plane Zero (RHPZ) transfer functions described by

\begin{equation}
	G_{\text{RHPZ},d}^\text{DER}(s) = e^{-T_{\text{delay,}d}^\text{DER} \cdot s} \cdot \frac{K_{d} - 0.25 \cdot K_{d} \cdot s}{(\frac{T_{\text{unc},d}^\text{DER}}{3.7348})^2 \cdot s^2 + \frac{T_{\text{unc},d}^\text{DER}}{0.9337} \cdot s + 1}. 
	\label{GRHPZ}
\end{equation}
\noindent
A RHPZ actuator reacts at first in the wrong direction and then follows the request of the correction variable. Therefore, a RHPZ actuator is a worst case for the control loop. Eq. \ref{GPT2} and Eq. \ref{GRHPZ} are chosen so that $G_{\text{PT1},d}^\text{DER}(s)$, $G_{\text{PT2},d}^\text{DER}(s)$, and $G_{\text{RHPZ},d}^\text{DER}(s)$ have the same rise time for each DER $d$.
 
In the following, $b$ is increased as well as decreased until the transfer function $G_{A_n,\text{unc}}^\text{CL}(s)$ becomes unstable for at least one combination of DERs $A_n$, starting with $b = 1$. The step response of G is used to check whether G is stable. During this, b is noted, from which the rise time and the settling time for at least one $A_n$ have doubled. The results of the uncertainty analysis are shown in Table \ref{Uncertainties}. 
\begin{table}[htbp]
	\caption{Influence of uncertainties}
	\begin{center}
		\begin{tabular}{|c|c|c|c|c|c|c|}
			\hline
			\textbf{Model } &\multicolumn{2}{|c|}{\textbf{Stability limit}} &\multicolumn{2}{|c|}{\textbf{2$\cdot$Rise Time}} &\multicolumn{2}{|c|}{\textbf{2$\cdot$Settling Time}}  \\
			\cline{2-7} 
			\rule{0pt}{8pt}\textbf{Type} & \textbf{\textit{b}$_{\text{min}}$} & \textbf{\textit{b}$_{\text{max}}$}& \textbf{\textit{b}$_{\text{min}}$}& \textbf{\textit{b}$_{\text{max}}$}& \textbf{\textit{b}$_{\text{min}}$}& \textbf{\textit{b}$_{\text{max}}$}  \\
			\hline
			PT1 & 0.0136 & $\infty$ & 0.0109 & 7.66 & 0.883 & 1.09  \\
			\hline
			PT2 & 0.0154 & 3.97 & 0.01 & 4.78 & 1 & 1  \\
			\hline
			RHPZ & 0.496 & 13.6 & 0.244 & 8.29 & 0.73 & 1   \\
			\hline
		\end{tabular}
		\label{Uncertainties}
	\end{center}
\end{table}

If the actuators correspond to a first-order transfer function (model type PT1) as assumed in the reference model, $G_{A_n,\text{unc}}^\text{CL}(s)$ will not become unstable by increasing $b$. This does not apply if $b$ is reduced or the model type differs from a first-order transfer function. Only if $T_{\text{unc},d}^\text{DER}$ is smaller than 1.36\,\% of $T_{d}^\text{DER}$ by using the model type PT1, the control loop $G_{A_n,\text{unc}}^\text{CL}(s)$ becomes unstable for at least one combination $A_n$. If all actuators correspond to model type RHPZ, $G_{A_n,\text{unc}}^\text{CL}(s)$ becomes already unstable at approximately a halving of $T_{d}^\text{DER}$. In contrast, the control loop $G_{A_n,\text{unc}}^\text{CL}(s)$ with model type RHPZ is comparatively robust against increasing $b$. In the investigations with increased $b$, the instability of $G_{A_n,\text{unc}}^\text{CL}(s)$ occurs earliest using model type PT2 with almost a fourfold increase of $T_{d}^\text{DER}$. 
The rise time is relatively robust to changes in $b$. In contrast, the settling time doubles by minor changes. The high sensitivity of the settling time has the following reason. Even slight deviations from the reference model can cause overshooting and oscillations that can significantly increase the time until the output response of $G_{A_n,\text{unc}}^\text{CL}(s)$ remains within the error band of the settling time. Summarized, the results in Table \ref{Uncertainties} illustrate the robustness against model uncertainties of the approach presented in Section \ref{CP}.   

\subsection{Exemplary Time-Domain Simulation}
\label{timdom}

Finally, simulation results of the detailed model are shown to demonstrate a more application-oriented use of the adaptive control system. The Simbench study case with the acronym lW from \cite{simbench2021grid} is simulated in the following investigation \cite{simbench2021grid}. This study case is characterized by low load and very high wind generation. At the interconnection point to the HV grid, energy is transferred from the MV grid to the HV grid. 

\begin{figure*}[h]
	\centering
	\includegraphics[scale=0.40]{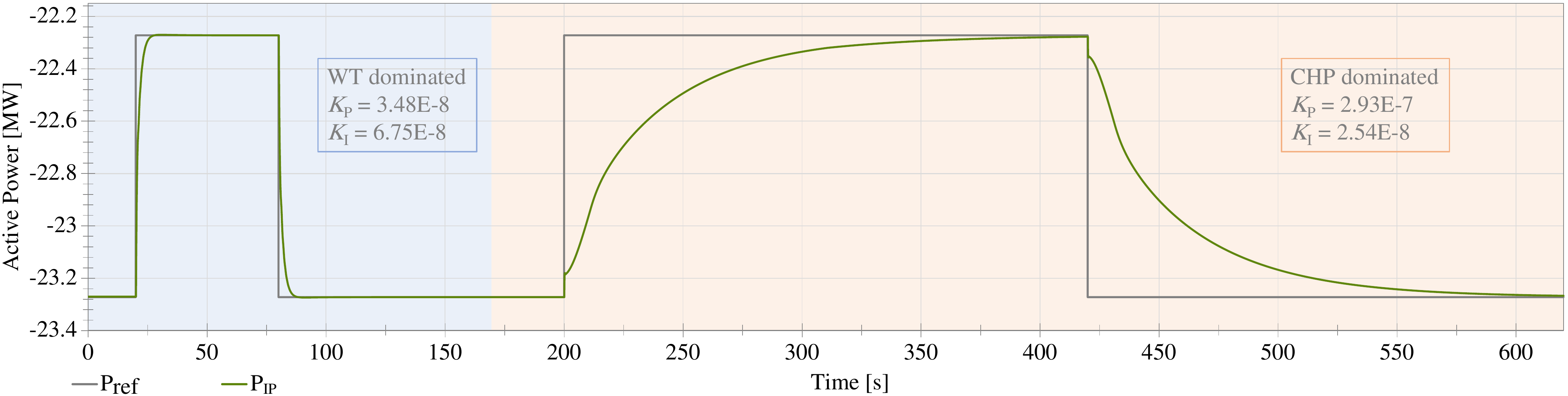}
	\caption{Exemplary time-domain simulation with WT dominated and CHP dominated set of actuators $A$}
	\label{hoecker}
\end{figure*}

The control system is excited by changes in the reference value $P_\text{ref}$. At time $t$ = 20\,s and $t$ = 200\,s, $P_\text{ref}$ is increased by a step of 1\,MW and reduced by 1\,MW a short time later. In the time span from 0\,s to 170\,s, all DERs are available to the CPFC. From 170\,s, the CPFC can no longer use the six WTs (compare Table \ref{DERpara}), so the adaptive controller adjusts the control parameters. Fig. \ref{hoecker} shows the interconnection power flow $P_\text{IP}$ to the 110\,kV grid and the corresponding reference value $P_\text{ref}$.

For all $P_\text{ref}$ jumps, the CPFC can completely compensate the deviation between $P_\text{ref}$ and $P_\text{IP}$. Despite the differences between the reference models and the detailed models, no oscillations occur after the four $P_\text{ref}$ jumps. For the first two $P_\text{ref}$ jumps the rise time and settling time lie within a few seconds. In contrast, for the $P_\text{ref}$ jumps at $t = 200\,\text{s}$ and $t = 420\,\text{s}$, the control system's rise and settling time are significantly higher than after the first two $P_\text{ref}$ jumps. This is caused by the unavailability of all WTs for the control system. The inert CHP plants dominate the performance of the control loop in this time span. Because of the presented control structure extension in Fig. \ref{stucturecpswitch}, the control parameter switching at t = 170\,s does not lead to deviations between $P_\text{IP}$ and $P_\text{ref}$. Furthermore, simulations of other study cases with different combinations of DERs have shown results comparable to Fig. \ref{hoecker}. Summarized, the exemplary simulation results in Fig. \ref{hoecker} demonstrate the operation of the adaptive control system developed in this work.         

\section{Conclusion and Outlook}
\label{conclusion}

This paper presents a method for designing an adaptive control system based on gain scheduling. For this purpose, a method for parameter tuning is developed. The simulation results show that the resulting adaptive control system operates with increased performance compared to a static control system. The performance depends significantly on the DERs available to the control system. In addition, the adaptive control system is proven to be highly robust, which is reflected in high margins and pointed out by uncertainty analysis. Even considering different dynamic behavior of the actuators, the CPFC shows high robustness. Clear stability limits are shown, which must be taken into account in the design and operation of the adaptive control system.

In future work, the method for designing the adaptive control system should also be examined based on further grid models in addition to the Simbench 20\,kV grid. The stability margins required and the uncertainties to be considered must be evaluated through laboratory and field studies. At first, Power Hardware-in-the-Loop simulations in the laboratory are recommended. Through such laboratory tests, the influence of real actuators on the CPFC can be investigated.

\section*{Acknowledgment}

The authors gratefully acknowledge funding by the German Federal Ministry of Education and Research (BMBF) within the Kopernikus Project ENSURE ‘New ENergy grid StructURes for the German Energiewende’ (funding number 03SFK1V0-2).

\bibliographystyle{IEEEtran}
\bibliography{references}

\end{document}